\documentclass[12pt,a4paper]{article}
\pdfoutput=1

\usepackage{microtype}
\usepackage{subfig} 
\usepackage{rotating}
\usepackage{fp,calc}
\usepackage{appendix}
\usepackage{placeins}
                                                         
\usepackage{lineno}  

\usepackage{graphicx}  
\usepackage{color}
\usepackage{colortbl}

\newcommand*\patchAmsMathEnvironmentForLineno[1]{%
\expandafter\let\csname old#1\expandafter\endcsname\csname #1\endcsname
\expandafter\let\csname oldend#1\expandafter\endcsname\csname
end#1\endcsname
 \renewenvironment{#1}%
   {\linenomath\csname old#1\endcsname}%
   {\csname oldend#1\endcsname\endlinenomath}%
}
\newcommand*\patchBothAmsMathEnvironmentsForLineno[1]{%
  \patchAmsMathEnvironmentForLineno{#1}%
  \patchAmsMathEnvironmentForLineno{#1*}%
}
\AtBeginDocument{%
\patchBothAmsMathEnvironmentsForLineno{equation}%
\patchBothAmsMathEnvironmentsForLineno{align}%
\patchBothAmsMathEnvironmentsForLineno{flalign}%
\patchBothAmsMathEnvironmentsForLineno{alignat}%
\patchBothAmsMathEnvironmentsForLineno{gather}%
\patchBothAmsMathEnvironmentsForLineno{multline}%
}

\usepackage{xspace}  
\usepackage{rotating}
\usepackage{multirow}

\usepackage{amsmath}
\usepackage{ifthen} 
\usepackage{hyperref}    
\usepackage[all]{hypcap} 
\newboolean{pdflatex}
\setboolean{pdflatex}{true} 
\newboolean{uprightparticles}
\setboolean{uprightparticles}{false} 
\usepackage{amssymb}
\usepackage{amsfonts}
\usepackage{upgreek}
\usepackage{rotating}
\usepackage{multirow}
\usepackage{ctable}
\graphicspath{{./figs/}} 

\newboolean{articletitles}
\setboolean{articletitles}{true}

\def\lhcb {\mbox{LHCb}\xspace}
\def\ux85 {\mbox{UX85}\xspace}

\ifthenelse{\boolean{uprightparticles}}%
{

 \def\PDelta      {\ensuremath{\Delta}\xspace}                 
 \def\PXi      {\ensuremath{\Xi}\xspace}                 
 \def\PLambda      {\ensuremath{\Lambda}\xspace}                 
 \def\PSigma      {\ensuremath{\Sigma}\xspace}                 
 \def\POmega      {\ensuremath{\Omega}\xspace}                 
 \def\PUpsilon      {\ensuremath{\Upsilon}\xspace}                 
 
 \def\PB      {\ensuremath{\mathrm{B}}\xspace}                 
                  
 \def\PD      {\ensuremath{\mathrm{D}}\xspace}

 \def\PK      {\ensuremath{\mathrm{K}}\xspace}

 \def\Pe      {\ensuremath{\mathrm{e}}\xspace}

 \def\Pi      {\ensuremath{\mathrm{i}}\xspace}

}
{

 \mathchardef\PDelta="7101
 \mathchardef\PXi="7104
 \mathchardef\PLambda="7103
 \mathchardef\PSigma="7106
 \mathchardef\POmega="710A
 \mathchardef\PUpsilon="7107
                  
 \def\PB      {\ensuremath{B}\xspace}                 
                  
 \def\PD      {\ensuremath{D}\xspace}

 \def\PK      {\ensuremath{K}\xspace}

 \def\Pe      {\ensuremath{e}\xspace}

 \def\Pi      {\ensuremath{i}\xspace}

}

\def\en         {\ensuremath{\Pe^-}\xspace}   
\def\ep         {\ensuremath{\Pe^+}\xspace}

\def\kaon  {\ensuremath{\PK}\xspace}
  \def\Kbar  {\kern 0.2em\overline{\kern -0.2em \PK}{}\xspace}

\def\Kz    {\ensuremath{\kaon^0}\xspace}
\def\Kzb   {\ensuremath{\Kbar^0}\xspace}
\def\KzKzb {\ensuremath{\Kz \kern -0.16em \Kzb}\xspace}
\def\Kp    {\ensuremath{\kaon^+}\xspace}
\def\Km    {\ensuremath{\kaon^-}\xspace}

\def\KpKm  {\ensuremath{\Kp \kern -0.16em \Km}\xspace}
\def\KS    {\ensuremath{\kaon^0_{\rm\scriptscriptstyle S}}\xspace} 
\def\KL    {\ensuremath{\kaon^0_{\rm\scriptscriptstyle L}}\xspace}

  \def\Dbar    {\kern 0.2em\overline{\kern -0.2em \PD}{}\xspace}
\def\D       {\ensuremath{\PD}\xspace}

\def\Dz      {\ensuremath{\D^0}\xspace}
\def\Dzb     {\ensuremath{\Dbar^0}\xspace}
\def\DzDzb   {\ensuremath{\Dz {\kern -0.16em \Dzb}}\xspace}
\def\Dp      {\ensuremath{\D^+}\xspace}
\def\Dm      {\ensuremath{\D^-}\xspace}

\def\DpDm    {\ensuremath{\Dp {\kern -0.16em \Dm}}\xspace}

  \def\Bbar    {\kern 0.18em\overline{\kern -0.18em \PB}{}\xspace}

  \def\Y#1S{\ensuremath{\PUpsilon{(#1S)}}\xspace}

\def\Lbar {\ensuremath{\kern 0.1em\overline{\kern -0.1em\PLambda}}\xspace}

\newcommand{\decay}[2]{\ensuremath{#1\!\to #2}\xspace}         

\def\to                 {\ensuremath{\rightarrow}\xspace}

\def\CP                {\ensuremath{C\!P}\xspace}

\def\AT#1     {\ensuremath{A_{\mathrm{T}}^{#1}}\xspace}           

\def\C#1      {\ensuremath{\mathcal{C}_{#1}}\xspace}                       
\def\Cp#1     {\ensuremath{\mathcal{C}_{#1}^{'}}\xspace}                    
\def\Ceff#1   {\ensuremath{\mathcal{C}_{#1}^{\mathrm{(eff)}}}\xspace}        
\def\Cpeff#1  {\ensuremath{\mathcal{C}_{#1}^{'\mathrm{(eff)}}}\xspace}       
\def\Ope#1    {\ensuremath{\mathcal{O}_{#1}}\xspace}                       
\def\Opep#1   {\ensuremath{\mathcal{O}_{#1}^{'}}\xspace}                    

\newcommand{\tev}{\ensuremath{\mathrm{\,Te\kern -0.1em V}}\xspace}
\newcommand{\gev}{\ensuremath{\mathrm{\,Ge\kern -0.1em V}}\xspace}
\newcommand{\mev}{\ensuremath{\mathrm{\,Me\kern -0.1em V}}\xspace}
\newcommand{\kev}{\ensuremath{\mathrm{\,ke\kern -0.1em V}}\xspace}
\newcommand{\ev}{\ensuremath{\mathrm{\,e\kern -0.1em V}}\xspace}
\newcommand{\gevc}{\ensuremath{{\mathrm{\,Ge\kern -0.1em V\!/}c}}\xspace}
\newcommand{\mevc}{\ensuremath{{\mathrm{\,Me\kern -0.1em V\!/}c}}\xspace}
\newcommand{\gevcc}{\ensuremath{{\mathrm{\,Ge\kern -0.1em V\!/}c^2}}\xspace}
\newcommand{\gevgevcccc}{\ensuremath{{\mathrm{\,Ge\kern -0.1em V^2\!/}c^4}}\xspace}
\newcommand{\mevcc}{\ensuremath{{\mathrm{\,Me\kern -0.1em V\!/}c^2}}\xspace}

\def\mm   {\ensuremath{\rm \,mm}\xspace}

\def\mum  {\ensuremath{\,\upmu\rm m}\xspace}

\def\invfb   {\ensuremath{\mbox{\,fb}^{-1}}\xspace}

\def\ps   {\ensuremath{{\rm \,ps}}\xspace}

\newcommand{\chisq}{\ensuremath{\chi^2}\xspace}

\def\gsim{{~\raise.15em\hbox{$>$}\kern-.85em
          \lower.35em\hbox{$\sim$}~}\xspace}
\def\lsim{{~\raise.15em\hbox{$<$}\kern-.85em
          \lower.35em\hbox{$\sim$}~}\xspace}

\def\pt         {\mbox{$p_{\rm T}$}\xspace}

\def\evtgen     {\mbox{\textsc{EvtGen}}\xspace}
\def\pythia     {\mbox{\textsc{Pythia}}\xspace}

\def\geant      {\mbox{\textsc{Geant4}}\xspace}

\def\photos     {\mbox{\textsc{Photos}}\xspace}

\def\tell1  {TELL1\xspace}
\def\ukl1   {UKL1\xspace}

\newcommand{\ie}{\mbox{\itshape i.e.}\xspace}

\newcommand{\CL}{C.L.\ }
\newcommand{\CLsb}{\ensuremath{\textrm{CL}_{\textrm{s+b}}}\xspace}
\newcommand{\CLs}{\ensuremath{\textrm{CL}_{\textrm{s}}}\xspace}
\newcommand{\CLb}{\ensuremath{\textrm{CL}_{\textrm{b}}}\xspace}

\newcommand{\swave}{{S-wave}\xspace}
\newcommand{\pwave}{{P-wave}\xspace}

\newcommand{\Ks}{\KS}
\newcommand{\Kl}{\KL}

\newcommand{\Ksmumu}{\ensuremath{\Ks\to\mu^+\mu^-}\xspace}
\newcommand{\Kmumu}{\ensuremath{K^0_{\rm\scriptscriptstyle L,S}\to\mu^+\mu^-}\xspace}

\newcommand{\Klmumu}{\ensuremath{\Kl\to\mu^+\mu^-}\xspace}
\newcommand{\Kspipi}{\ensuremath{\Ks\to\pi^+\pi^-}\xspace}

\newcommand{\Kspimunu}{\ensuremath{\Ks\to\pi^+\mu^-\bar{\nu}_\mu}\xspace}

\newcommand{\Jpsimumu}{\ensuremath{J/\psi\to \mu^+\mu^-}\xspace}
\newcommand{\Jpsi}{\ensuremath{J/\psi}\xspace}

\newcommand{\BRof}[1]{\ensuremath{{\cal B}(#1)}\xspace}
\newcommand{\IP}{\ensuremath{{\rm IP}}\xspace}

\newcommand{\etos}{\ensuremath{\epsilon^\text{TOS/SEL}}\xspace}

\newcommand{\etrig}{\ensuremath{\epsilon^\text{TRIG/SEL}}\xspace}
\newcommand{\epid}{\ensuremath{\epsilon^\text{PID}}\xspace}
\newcommand{\emb}{\ensuremath{s^\text{MB}}\xspace}

\newcommand{\eSelect}{\ensuremath{\epsilon^\text{SEL}}\xspace}

\newcommand{\figref}[1]{Fig.~\ref{#1}}
\newcommand{\tabref}[1]{Table~\ref{#1}}

\newcommand{\secref}[1]{Sect.~\ref{#1}}

\newcommand{\tabcaption}[1]{  
\vspace{-\abovecaptionskip} %
\caption[#1]{#1}
\vspace{\abovecaptionskip}
}

\usepackage{cite} 
\usepackage{mciteplus}
\begin{document}

\renewcommand{\thefootnote}{\fnsymbol{footnote}}
\setcounter{footnote}{1}



\begin{titlepage}
\pagenumbering{roman}

\vspace*{-1.5cm}
\centerline{\large EUROPEAN ORGANIZATION FOR NUCLEAR RESEARCH (CERN)}
\vspace*{1.5cm}
\hspace*{-0.5cm}
\begin{tabular*}{\linewidth}{lc@{\extracolsep{\fill}}r}
\ifthenelse{\boolean{pdflatex}}
{\vspace*{-2.7cm}\mbox{\!\!\!\includegraphics[width=.14\textwidth]{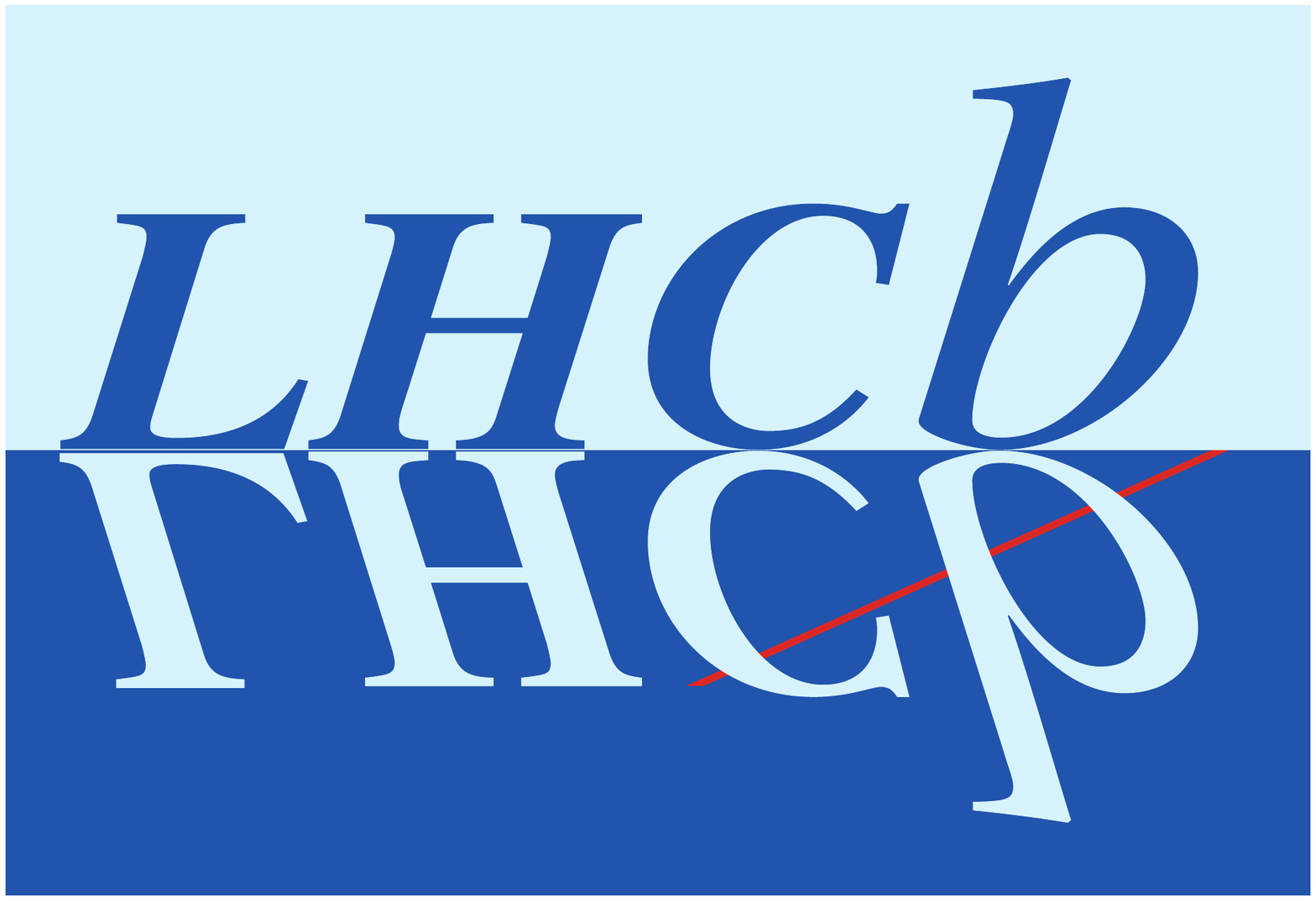}} & &}%
{\vspace*{-1.2cm}\mbox{\!\!\!\includegraphics[width=.12\textwidth]{lhcb-logo.eps}} & &}%
\\
 & & CERN-PH-EP-2012-267 \\  
 & & LHCb-PAPER-2012-023 \\  
 & & 18 September 2012 \\ 
 & & \\
\end{tabular*}
\vspace*{3.5cm}

{\bf\boldmath\huge
\begin{center}
  Search for the rare decay $K_{\rm\scriptscriptstyle S}^0\rightarrow\mu^{+}\mu^{-}$

\end{center}
}

\vspace*{2.0cm}

\begin{center}
The LHCb collaboration\footnote{Authors are listed on the following pages.}
\end{center}

\vspace{\fill}

\begin{abstract}
\noindent
A search for the decay $K_{\rm\scriptscriptstyle S}^0\rightarrow\mu^{+}\mu^{-}$ is performed, based on a data sample of 1.0\,fb$^{-1}$ of $pp$ collisions at $\sqrt{s}$ = 7\,TeV 
collected by the LHCb experiment at the Large Hadron Collider. 
The observed number of candidates is consistent with the background-only hypothesis, yielding an
upper limit of $\mathcal{B}(K_{\rm\scriptscriptstyle S}^0\rightarrow\mu^{+}\mu^{-}) < 11 (9)\times10^{-9}$ at 95 (90)$\%$ confidence level.
This limit is a factor of thirty below the previous
measurement.
\end{abstract}

\vspace*{1.0cm}

\begin{center}
  Published in the Journal of High Energy Physics
\end{center}

\vspace{\fill}

\end{titlepage}


\newpage
\setcounter{page}{2}
\mbox{~}
\newpage

\centerline{\large\bf LHCb collaboration}
\begin{flushleft}
\small
R.~Aaij$^{38}$, 
C.~Abellan~Beteta$^{33,n}$, 
A.~Adametz$^{11}$, 
B.~Adeva$^{34}$, 
M.~Adinolfi$^{43}$, 
C.~Adrover$^{6}$, 
A.~Affolder$^{49}$, 
Z.~Ajaltouni$^{5}$, 
J.~Albrecht$^{35}$, 
F.~Alessio$^{35}$
% ...

{\footnotesize \it
$ ^{1}$Centro Brasileiro de Pesquisas F\'{i}sicas (CBPF), Rio de Janeiro, Brazil\\
% ...
$ ^{5}$Clermont Universit\'{e}, Universit\'{e} Blaise Pascal, CNRS/IN2P3, LPC, Clermont-Ferrand, France\\
$ ^{6}$CPPM, Aix-Marseille Universit\'{e}, CNRS/IN2P3, Marseille, France\\
% ...
$ ^{11}$Physikalisches Institut, Ruprecht-Karls-Universit\"{a}t Heidelberg, Heidelberg, Germany\\
% ...
$ ^{33}$Universitat de Barcelona, Barcelona, Spain\\
$ ^{34}$Universidad de Santiago de Compostela, Santiago de Compostela, Spain\\
$ ^{35}$European Organization for Nuclear Research (CERN), Geneva, Switzerland\\
% ...
$ ^{38}$Nikhef National Institute for Subatomic Physics, Amsterdam, The Netherlands\\
% ...
$ ^{43}$H.H. Wills Physics Laboratory, University of Bristol, Bristol, United Kingdom\\
% ...
$ ^{49}$Oliver Lodge Laboratory, University of Liverpool, Liverpool, United Kingdom\\
% ...
$ ^{n}$LIFAELS, La Salle, Universitat Ramon Llull, Barcelona, Spain\\
% ...
}
\end{flushleft}

\cleardoublepage


\renewcommand{\thefootnote}{\arabic{footnote}}
\setcounter{footnote}{0}


%


\pagestyle{plain} 
\setcounter{page}{1}
\pagenumbering{arabic}


\section{Introduction}
\label{sec:intro}

The decay \Ksmumu is a Flavour Changing Neutral Current (FCNC) transition that has not yet
been observed. This decay is suppressed in the Standard Model (SM), with an expected 
branching fraction~\cite{Ecker:1991ru,Isidori:2003ts}
\begin{equation}
\BRof\Ksmumu = (5.0 \pm 1.5) \times 10^{-12},  \nonumber
\end{equation}
\noindent while the current experimental upper limit is $3.2\times10^{-7}$ at 90$\%$ confidence level (C.L.) \cite{Jack}.

Although the dimuon decay of the \Kl meson is known to be \mbox{$\BRof\Klmumu = ( 6.84 \pm 0.11 ) \times 10^{-9}$}~\cite{PDG}, in agreement with the SM, effects of new particles can still be observed
in \Ksmumu decays. In the most general case, the decay width of \Kmumu can be written as~\cite{Gino2}
\begin{equation}
\Gamma(\Kmumu) = \frac{m_K}{8\pi}\sqrt{1-\left(\frac{2 m_{\mu}}{m_K}\right)^2} \left[|A|^2 + \left(1-\left(\frac{2 m_{\mu}}{m_K}\right)^2\right)|B|^2\right],
\end{equation}
\noindent where $A$ is an \swave amplitude and $B$ a \pwave amplitude. These two amplitudes have opposite \CP eigenvalues, and in absence of
\CP violation (\Ks $= K_{1}^{0}$, \Kl $=K_{2}^{0}$), \Kl decays would be generated only by $A$ while \Ks decays would be generated only by $B$.  
The decay width $\Gamma(\Klmumu)$ receives long-distance\footnote{The long-distance scales correspond to masses below that of the $c$ quark, while short-distance scales correspond to masses of the $c$ quark and above.}  contributions to $A$ from intermediate two-photon states, as well as  short distance contributions to the real part of $A$. In any model with the same basis of
effective FCNC operators as the SM, the contributions from $B$ can be neglected for \BRof\Klmumu. The decay width of \Ksmumu depends on the 
imaginary part of the short-distance contributions to $A$  and on the long-distance contributions to $B$ generated by intermediate two-photon states.
Therefore, the measurement of \BRof\Klmumu in agreement with the SM does not necessarily imply that \BRof\Ksmumu has to agree with the SM.
Contributions up to one order of magnitude above the SM expectation
are allowed~\cite{Isidori:2003ts}; enhancements of the branching fraction above $10^{-10}$ are less likely.
The study of \Ksmumu has been suggested as a possible way to look for new light scalars \cite{Ecker:1991ru}.

In addition, bounds on the upper limit of \BRof\Ksmumu close to $10^{-11}$ could be very useful to discriminate among scenarios beyond the SM if other modes, such
as $K^{+}\to\pi^{+}\nu\bar{\nu}$ (charge conjugation is implied throughout this paper),
were to indicate a non-standard enhancement of the $s\to d \ell\bar{\ell}$ transition \cite{Isidori:2003ts}.
The KLOE collaboration has searched for the related decay $\Ks \to \ep\en$, which is affected by a larger helicity suppression than the muonic mode, and set an upper limit on the branching fraction $\BRof{\Ks \to \ep\en}<9\times10^{-9}$ 
at 90\% confidence level \cite{Ambrosino:2008zi}.

The LHC produces $\sim10^{13}$ \Ks per \invfb inside the LHCb acceptance. 
In this paper, a search for \Ksmumu is presented using 1.0 fb$^{-1}$ of $pp$ collisions at $\sqrt{s}$ = 7 TeV collected by LHCb in 2011. 
Dimuon candidates 
are classified in bins of a multivariate discriminant, and compared
to background and signal expectations.
The background present in the signal region is a combination of combinatorial background
and \Kspipi decays in which both pions are misidentified as muons. 
The number of expected signal candidates for a given branching fraction hypothesis is obtained by normalising to the measured \Kspipi rate.
The results obtained by the
measurements in different bins are combined, and a limit is set using the \CLs method~\cite{Read_02,Junk_99}.
The data in the signal region were only analysed once the full analysis strategy was defined, including the selection, the binning and the evaluation 
of systematic uncertainties.

The LHCb apparatus, and the aspects of the trigger relevant for this analysis are presented in \secref{sec:Detector}. 
Section \ref{sec:mva} is devoted to the full signal selection and to the definition of the multivariate method used as the main discriminant. 
In \secref{sec:backgrounds} the different backgrounds for \Ksmumu decay are described, as well as the expected background in the signal region.
The normalisation, required to convert the number of \Ksmumu candidates to the branching fraction, is detailed in \secref{sec:normalisation}. The
systematic uncertainties are described in \secref{sec:systematics}.
The limit setting procedure,
together with the corresponding expected and observed limits, is presented in \secref{sec:results}, and conclusions are drawn in \secref{sec:conclusions}. %

\section{Experimental setup}
\label{sec:Detector}

The \lhcb detector~\cite{Alves:2008zz} is a single-arm forward
spectrometer covering the \mbox{pseudorapidity} range $2<\eta <5$, designed
for the study of particles containing $b$ or $c$ quarks. The
detector includes a high precision tracking system consisting of a
silicon-strip vertex detector (VELO) surrounding the $pp$ interaction region,
a large-area silicon-strip detector located upstream of a dipole
magnet with a bending power of about $4{\rm\,Tm}$, and three stations
of silicon-strip detectors and straw drift tubes placed
downstream. The combined tracking system has a momentum resolution
$\Delta p/p$ that varies from 0.4\% at 5\gevc to 0.6\% at 100\gevc,
and an impact parameter (\IP) resolution of 20\mum for tracks with high
transverse momentum (\pt) with respect to the beam direction. 
Charged hadrons are identified using two
ring-imaging Cherenkov detectors. Photon, electron and hadron
candidates are identified by a calorimeter system consisting of
scintillating-pad and \mbox{preshower} detectors, an electromagnetic
calorimeter and a hadronic calorimeter. Muons are identified by a 
system composed of alternating layers of iron and multiwire
proportional chambers. 

The trigger consists of a hardware stage, based
on information from the calorimeter and muon systems, followed by a
software stage which applies a full event reconstruction.
For this analysis, the events are first required to pass a hardware trigger
which selects at least one muon with  $\pt>1.5\gevc$.  
In the subsequent software trigger~\cite{muon_trigger}, at least one of the final
state tracks is required to be of good quality and to have $\pt>1.3\gevc$,
an \mbox{\IP $>0.5$\mm} and the $\chisq$
of the impact parameter (IP $\chisq$) above $200$.
The IP $\chisq$ is defined as the difference between the $\chisq$
of the proton-proton, $pp$, interaction point (primary vertex, PV) built with 
and without the considered track.
A prescale factor of two is applied to the lines triggered by the \Ksmumu candidates.
The \Ksmumu candidates responsible for the trigger of both the hardware and
software levels are called TOS (trigger on signal).

Events with a reconstructed \Ksmumu candidate can also be triggered independently of the signal candidate 
if some other combination of particles in the underlying event passes the trigger. 
Such candidates are called TIS (trigger independently of signal). 
The TIS and TOS categories are not exclusive as muons from both the \Ksmumu candidates and from the underlying event can pass the trigger.
There is overlap between the two, which 
allows the determination of trigger efficiencies from the data~\cite{TISTOS}.
Finally, minimum bias candidates triggered by a dedicated random trigger (MB) 
provide a negligible amount of \Ksmumu candidates. Instead they allow the selection of a sample of 
\Kspipi useful to understand the distributions that the signal would have in the case of
no trigger bias.

For the simulation, $pp$ collisions are generated using
\pythia~6.4~\cite{Sjostrand:2006za} with a specific \lhcb
configuration~\cite{LHCb-PROC-2010-056}.  Decays of hadronic particles
are described by \evtgen~\cite{Lange:2001uf} in which final state
radiation is generated using \photos~\cite{Golonka:2005pn}. The
interaction of the generated particles with the detector and its
response are implemented using the \geant
toolkit~\cite{Allison:2006ve, *Agostinelli:2002hh} as described in
Ref.~\cite{LHCb-PROC-2011-006}.

\section{Selection and multivariate classifier}
\label{sec:mva}

The \Ksmumu candidates are reconstructed requiring two tracks with opposite curvature with hits in the VELO and
in the tracking stations. About 40\% of the \Ks mesons with the two daughter tracks inside the LHCb acceptance
decay in the VELO detector. 
Those tracks are required to be of high quality 
($\chi^2 < 5$ per degree of freedom), to have an \IP $\chi^2$ greater than 100 and a distance of closest approach
of less than 0.3 \mm. The two tracks are required to be identified as muons \cite{muonid_note}. 
The reconstructed \Ksmumu candidates are required to have a proper decay
time greater than 8.9 \ps and to point to the
PV ($\IP(\Ks)<400$ \mum). The secondary vertex, SV, of the \Ksmumu candidate is required
to be downstream of the PV.
If more than one PV is reconstructed, the PV associated to the \Ks is the one that minimises its \IP $\chi^2$.
Furthermore, $\Lambda \to p \pi^-$ decays are vetoed via a requirement in the
Armenteros-Podolanski plane \cite{Armenteros}, by including cuts on the transverse momentum of the
daughter tracks with  respect to the \Ks flight direction and on their longitudinal momentum asymmetry.
The reconstructed \Ksmumu mass is required to be in the range [450,1500] \mevcc.

The \Kspipi decay is used as a control channel and is reconstructed and selected in the same
way as the signal candidates, with the exception of the particle identification requirements
on the daughter tracks and the mass range, which is requested to be between 400 and 600 \mevcc.

\begin{figure}
	\centering
	\includegraphics[width=0.6\textwidth]{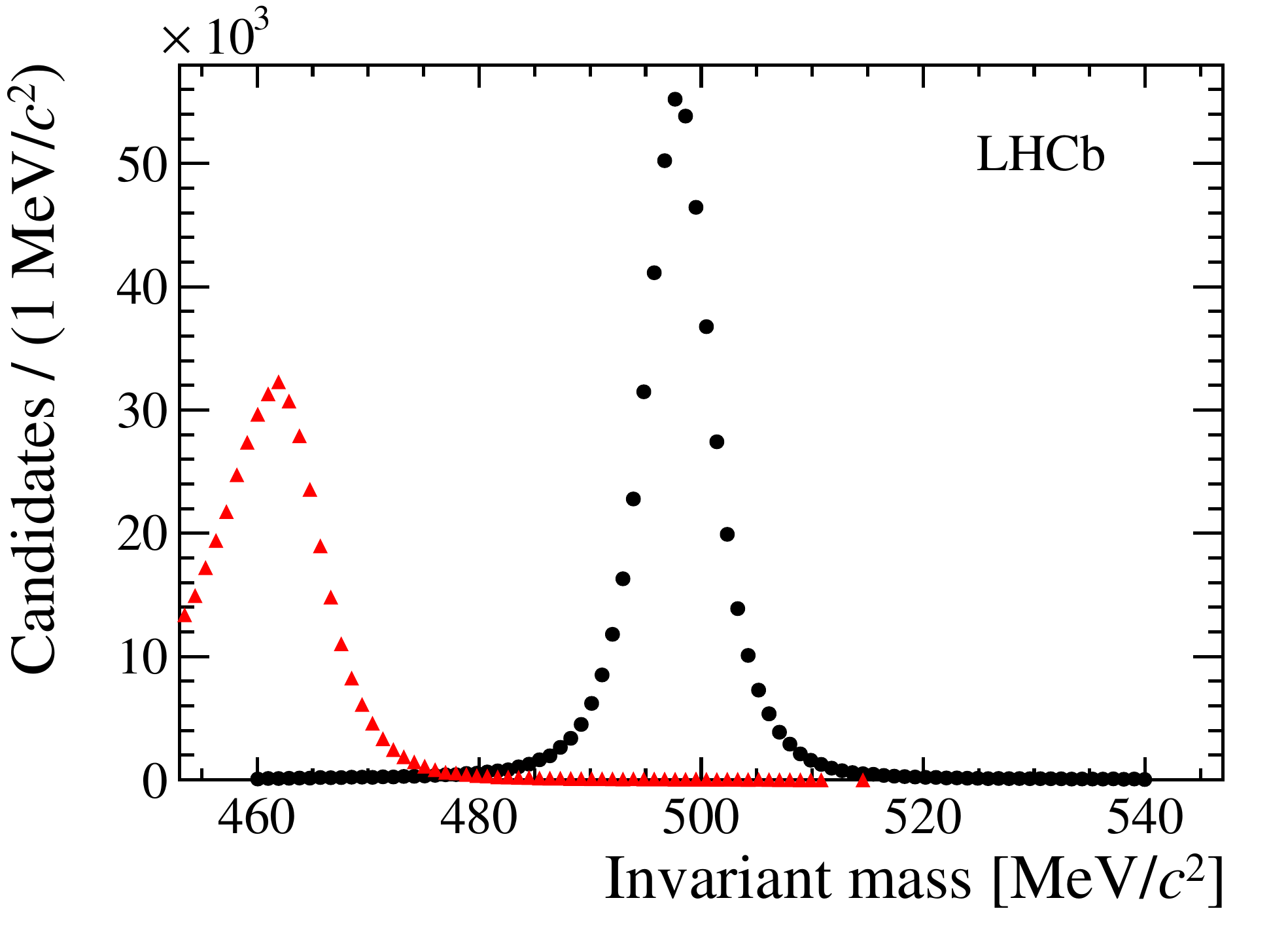}
	\caption{Mass spectrum for selected \Kspipi candidates in the MB sample. The black points correspond to the mass reconstructed under the $\pi\pi$ mass hypothesis for the daughters, while the red triangles correspond to the mass reconstructed under the $\mu\mu$ mass hypothesis.}
	\label{fig:kspipi_fit}
\end{figure}

Figure \ref{fig:kspipi_fit} shows the mass spectrum for selected \Kspipi candidates in the MB sample
after applying the set of cuts described above and in the $\pi\pi$ and $\mu\mu$ mass hypotheses:
the two mass peaks are separated by 40 \mevcc. 
This separation, combined with the LHCb mass resolution of about 4 \mevcc for 
such combinations of tracks, is used to discriminate the \Ksmumu signal from \Kspipi decays 
where both pions are misidentified as muons.

In order to further increase the background rejection, a boosted decision tree~(BDT)~\cite{Breiman} with the AdaBoost algorithm~\cite{AdaBoost} is used. 
The variables entering in the BDT discriminant are:
\begin{itemize}
\item{\it the decay time of the \Ks candidate}, computed using the distance between the SV and the PV,
and the reconstructed momentum of the \Ks candidate;
\item{\it the smallest muon IP $\chi^2$ of the two daughter tracks with respect to any 
of the PVs reconstructed in the event};
\item{\it the }\Ks\ {\it IP $\chi^2$ with respect to the PV};
\item{\it the distance of closest approach between the two daughter tracks}; 
\item{\it the secondary vertex $\chi^{2}$}, which adds complementary information with respect to the distance of closest approach of the tracks, as it uses information on the uncertainty of the vertex fit;
\item{\it the angle of the decay plane in the \Ks rest frame with respect to the \Ks flight direction,} which is isotropic for signal decays, but not necessarily for background candidates;

\item{\it variables used to discriminate against material interactions}, as further detailed below.
\end{itemize}
 
\begin{figure}[t]
\centering
\includegraphics[width=0.495\textwidth]{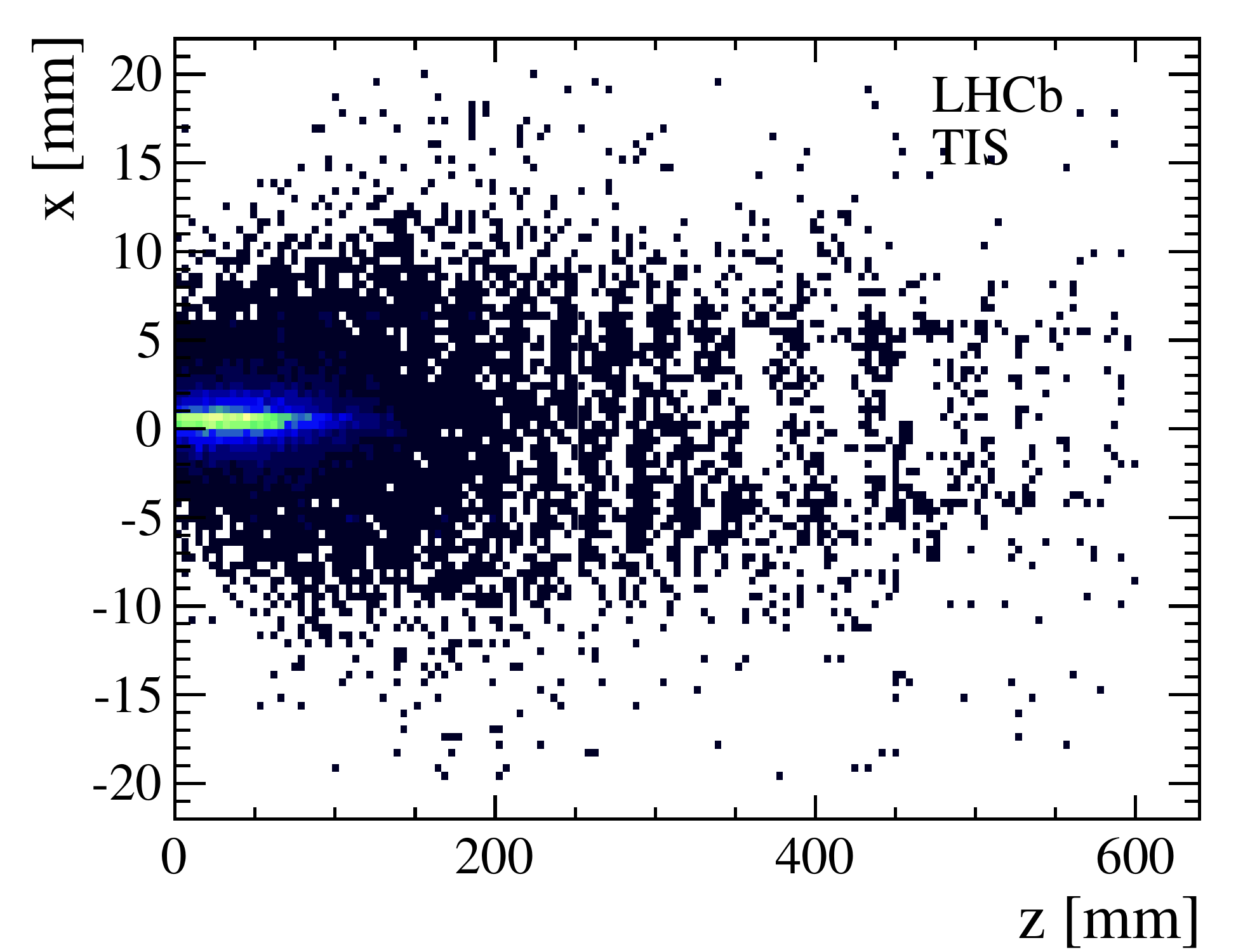}
\includegraphics[width=0.495\textwidth]{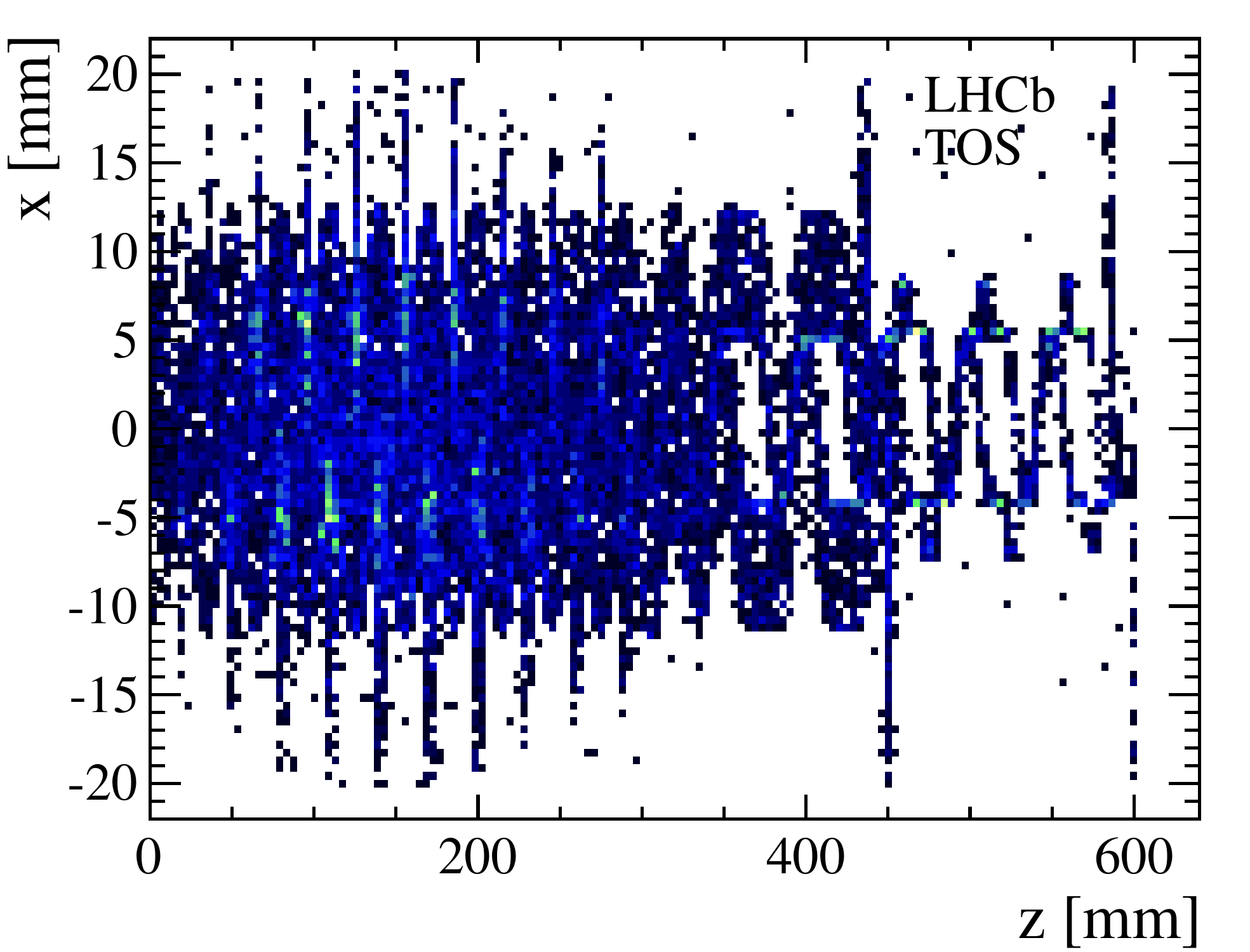}
\caption{Position in the  $x-z$ plane of the secondary vertices of the background candidates found in the high mass
sideband for (left) TIS candidates and (right) TOS candidates. The lighter coloured
areas correspond to higher density of points.}
\label{fig:VELO}
\end{figure}

An important source of background consists of muons resulting from interactions between the particles produced in the PV and the detector 
material in the region of the VELO.  The position of the SV of the background candidates from the \Ks mass sidebands in the $x-z$ plane is shown in \figref{fig:VELO}.
The structures observed correspond to the position of the material inside the VELO detector. 
To discriminate against this background, two different approaches are used for the TIS and TOS trigger categories, consisting of two different choices of variables for the BDT.

For the TOS category, two additional variables are included in the BDT, the \pt of the \Ks and a boolean \textit{matter veto} that uses the VELO geometry to assess whether a given decay vertex coincides with a point in the detector material or not.
Muons from material interactions have a harder \pt spectrum than muons from other background sources and hence are more likely to be selected by the trigger. The use of this variable in the BDT provides 50\% less background yield for the same signal efficiency than simply applying the veto as a selection cut.

For the TIS category, the coordinates of the position of the SV in the laboratory frame are used to deal with this background.
As the simultaneous use of the lifetime, \pt of the \Ks  meson, and the SV position allows the BDT to effectively compute 
the mass of the candidate, a fake signal peak could be artificially created out of the combinatorial background. Hence the \pt of the \Ks meson is not used in the TIS analysis.
This second approach provides a factor of two less background yield for the same signal efficiency than the {\it matter veto} (and \Ks \pt) for the TIS analysis, while, on the contrary, the {\it matter veto} boolean variable gives a factor of four less background yield for the same signal efficiency than the SV coordinates for the TOS analysis.

Because of these different approaches and to take into account the biases on the variable distributions introduced by the trigger, the data sample is split in two subsamples according to the TIS and TOS categories, for which BDT discriminants are optimised separately.
In the TOS analysis, the \Kspipi decays are required to have at least
one of the daughters with a \pt above 1.3 \gevc in order to minimise
the difference in the momentum distributions with respect to the
triggered \Ksmumu candidates.
The candidates that are simultaneously TIS and TOS are analysed only as TIS candidates to avoid counting them twice. Only one per mille of the TOS candidates overlap with TIS candidates.

In addition, the BDT discriminants for both trigger categories are defined and 
trained on data using \Kspipi candidates as signal sample and 
\Ksmumu candidates in the upper mass sideband as background sample.
For the background sample, the region above 1100 \mevcc (above the $\phi$ resonance) 
is used to define the BDT settings and the region between 504 and 1000 \mevcc 
to train the BDT algorithm chosen.
For the signal sample, the \Kspipi TIS events are used to train the BDT for the TIS category,
while \Kspipi decays with both pions misidentified as muons and passing 
the same trigger requirements as the \Ksmumu signal are used for the TOS category.
In order to minimise the differences between misidentified \Kspipi events and \Ksmumu decays,
tight muon identification requirements (including cuts in the quality of the tracks or in the number
of muon hits shared by different tracks) are applied to the \Kspipi sample. These tight requirements
are chosen such that the efficiency of the trigger in the \Kspipi simulated decays is the same as in the
\Ksmumu simulated decays.

In addition, the TOS and TIS categories are further split in two equal-sized subsamples, corresponding to the first and second halves of the data taking period.
This procedure prevents possible biases related to the use of the same events
in the mass sidebands both to train the BDT discriminant and to evaluate the background in
the signal region, while making maximal use of the available data both for BDT training and
background evaluation. Thus, in total, four different samples are defined (two subsamples 
for the TIS trigger category and two subsamples for the TOS trigger category) and combined as described 
in \secref{sec:results}.

Candidates with low values of the BDT response are not considered because of the large amount of background in that region. This requirement provides about $50\%$ signal efficiency and $99\%$ background rejection, depending on the sample.
The rest of the candidates are classified in ten bins of equal signal efficiency, \ie a total of forty bins are combined to get the \CLs limit.

\section{Background}
\label{sec:backgrounds}

\begin{figure}[t]
\centering
\includegraphics[width=0.495\textwidth]{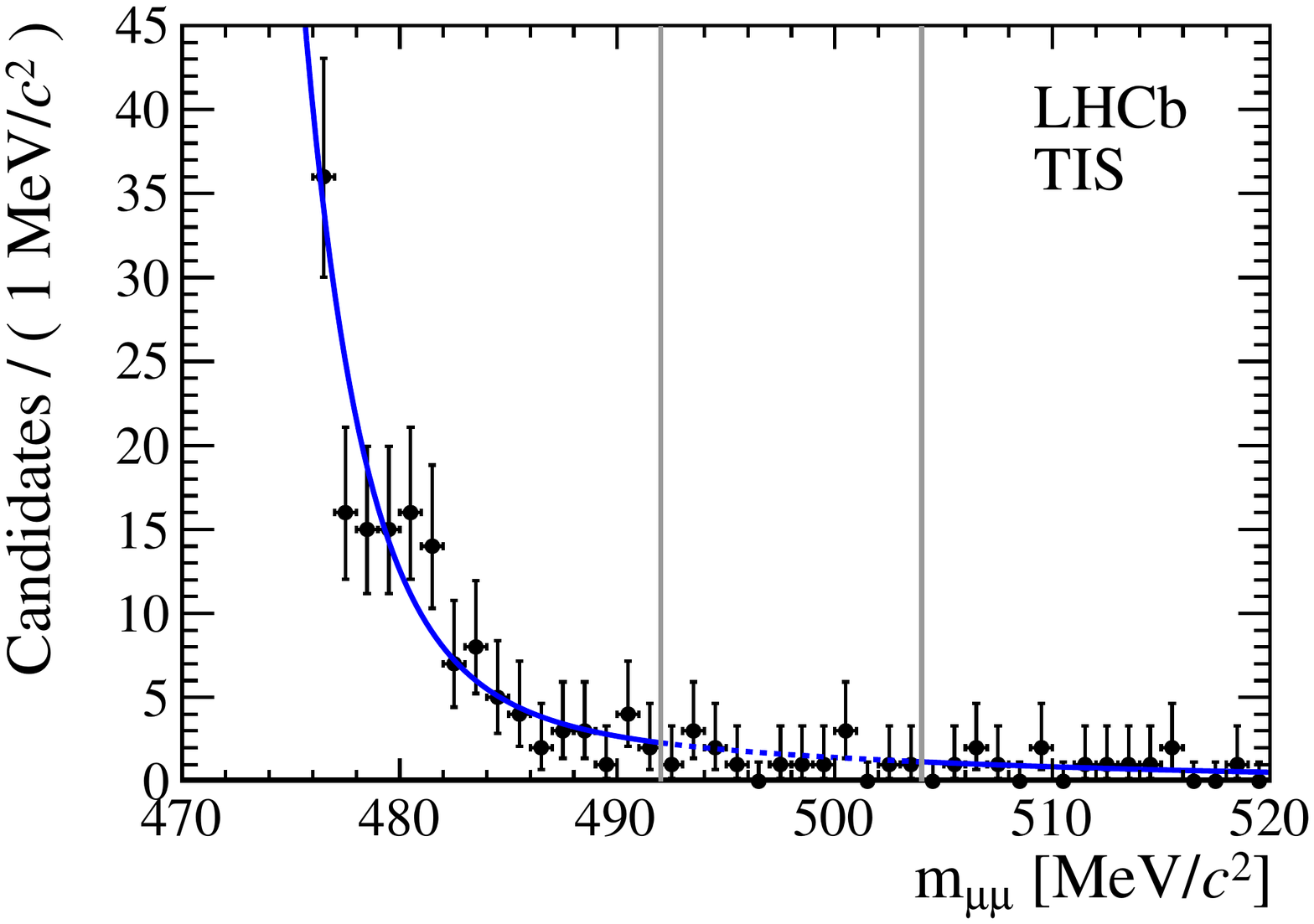}
\includegraphics[width=0.495\textwidth]{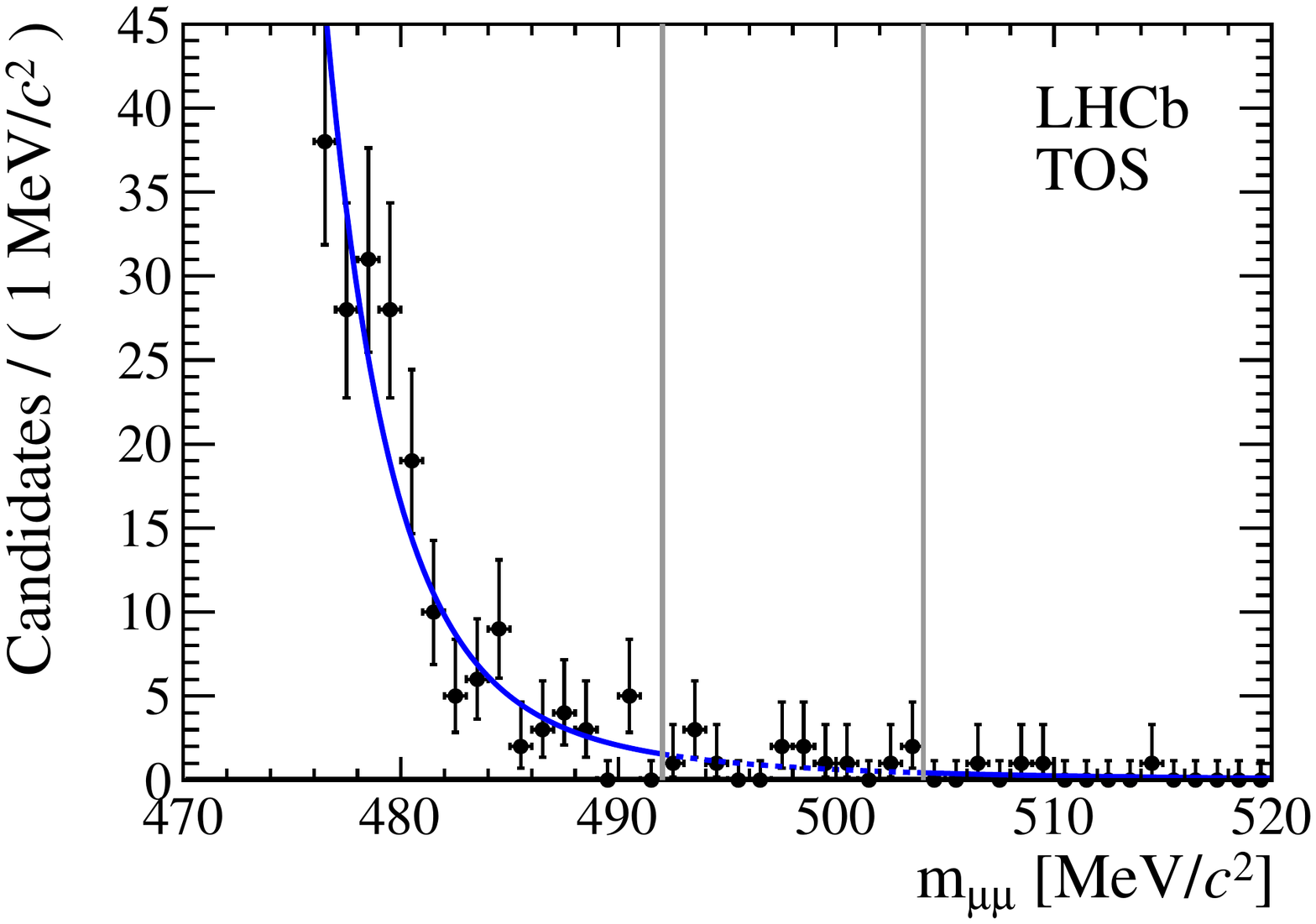}
\caption{Background model fitted to the data separated along (left) TIS and (right) TOS trigger categories. The vertical lines delimit the search window.}
\label{fig:combplots}
\end{figure}

The search region is defined as the mass range $[492,504]$\mevcc.
The background level is calibrated by interpolating the observed yield 
from mass sidebands ($[470,492]$ and $[504,600]$ \mevcc) to the signal region. This is done by means of an unbinned 
maximum likelihood fit in the sidebands, using a model with two components. 
The first component is a power law that describes the tail of \Kspipi decays where both pions are misidentified as muons;
this model has been checked to be appropriate using MC simulation.
The second component is an exponential function describing the combinatorial background.
As an illustration,~\figref{fig:combplots} shows the distribution of candidates for all BDT bins and for TIS and TOS samples, respectively. The expected total background yield in the most sensitive BDT bins of both samples ranges from 0 to 1 candidates.

Other sources of background, such as \Kspimunu, $\Ksmumu \gamma$, $\Kl\to\mu^{+}\mu^{-}\gamma$,
$\Kl\to\pi^+\mu^-\bar{\nu}_\mu$ and \Klmumu decays, are negligible for the current analysis.
In the case of \Klmumu and $\Klmumu \gamma$, the contributions have been evaluated
using the ratio
of the \Ks and \Kl lifetimes and the proper time acceptance measured in data 
with the \Kspipi decays.
The contributions of the other decay modes have been determined using MC simulated events.

\section{Normalisation}
\label{sec:normalisation}

A normalisation is required to translate the number of \Ksmumu signal decays into a branching fraction measurement. Two normalisations are determined independently for TIS and TOS candidates.
The \BRof\Ksmumu is computed using 
\begin{equation}
\frac{\BRof\Ksmumu}{\BRof\Kspipi} =
\frac{\epsilon_{\pi\pi}}{\epsilon_{\mu\mu}} \frac{N_{\Ksmumu}}{N_{\Kspipi}},
\end{equation}
where, in a given BDT bin, $N_{\Ksmumu}$ is the observed number of signal decays, 
$N_{\Kspipi}$ the number of \Kspipi decays,
and $\epsilon_{\pi\pi}/\epsilon_{\mu\mu}$ the ratio of the corresponding efficiencies.
The efficiencies are factorised as $\epsilon= \eSelect \epid \etrig$ where:
\begin{itemize}
\item \eSelect is the offline selection efficiency. It includes the geometrical acceptance, reconstruction and selection, i.e, it is the probability for a \Kspipi (\Ksmumu) 
decay generated in a $pp$ collision, to have been reconstructed and selected;
\item \epid is the efficiency of the muon identification for reconstructed and selected \Ksmumu signal decays;
\item $\etrig = N^\text{SEL\&PID\&TRIG}/N^\text{SEL\&PID}$, where TRIG denotes either the TIS or the TOS categories, is the trigger efficiency for decays that would be offline selected.
Under this definition, trigger efficiencies can be determined from data using the procedure described in Ref.~\cite{TISTOS}.
\end{itemize}

The ratio of reconstruction and selection efficiencies between \Ksmumu
and \Kspipi decays is evaluated in bins of \pt and rapidity of the \Ks meson
using simulated events
reweighted in order to reproduce the \Ks \pt and rapidity spectra measured
in data \cite{Raluca}.
 The reconstruction and selection efficiency for \Kspipi decays is between 60\% and 85\% (depending on  which point in the phase space a given event is from) of that of the \Ksmumu decays due to difference in the material
interactions of the pions compared to muons.

The factor \epid is evaluated in bins of the BDT (both for the TOS and TIS categories) by measuring the muon identification efficiency as a function of $p$ and \pt using 
calibration muons. The sample of calibration muons is obtained from a \Jpsimumu sample in which positive muon identification is required for only one of the tracks. 
The $p$ and \pt spectra of the pions from \Kspipi decays in a MB sample is later used to get the efficiency for \Ksmumu decays. The \epid efficiency is between $68\%$ and $82\%$ (depending on the BDT bin and the sample). It is measured with a precision between $1\%$ and $10\%$.
For the ratio of trigger efficiencies, different strategies are considered for the TIS and TOS samples.

For the TIS samples, the \Ksmumu yield is normalised to the \Kspipi TIS yield. In this case, the trigger efficiencies cancel in the ratio, because the probability
to trigger on the underlying event is independent of the decay mode of the \Ks meson. This cancellation is
verified in simulation.
The normalisation expression for TIS decays reads
\begin{equation}
\frac{\BRof\Ksmumu}{\BRof\Kspipi} = \frac{\eSelect_{\pi\pi}}{\eSelect_{\mu\mu}}   \frac{1}{\epid_{\mu\mu}} \frac{N_{\Ksmumu}^\text{TIS}}{N_{\Kspipi}^\text{TIS}}, 
\end{equation}
where $N^\text{TIS}_{\Ksmumu}$ and $N^\text{TIS}_{\Kspipi}$ are the number of TIS decays in a given BDT bin for signal and \Kspipi modes respectively. $N^\text{TIS}_{\Kspipi}$ is found to be around 9000 for every BDT bin.

For the TOS sample, the \Ksmumu yield is normalised to the \Kspipi yield from MB triggers.
The normalisation requires in this case an absolute determination of the TOS trigger efficiency for \Ksmumu, $\etos_{\mu\mu}$, as well as the knowledge of the average prescale factor of the MB trigger, \emb.
The absolute TOS trigger efficiency for the signal is computed 
using muons from \decay{B^+}{\Jpsi(\to\mu^+\mu^-)K^+} decays.\footnote{To avoid bias, it is required that another object be the origin of the trigger and not the muons alone, \ie the muons from this sample are TIS.} The $p$ and \pt spectra of the \decay{B^+}{\Jpsi(\to\mu^+\mu^-)K^+} muons are reweighted in order to match those of pions from the \Kspipi decays. Trigger unbiased $p$ and \pt spectra of the \Kspipi
decays can be obtained from the MB sample. The TOS efficiency is found to be at the level of 20\% for all BDT bins.
The normalisation expression for TOS decays reads
\begin{equation}
\frac{\BRof\Ksmumu}{\BRof\Kspipi} = \frac{\eSelect_{\pi\pi}}{\eSelect_{\mu\mu}} \frac{1}{\epid_{\mu\mu}}  \frac{\emb}{\etos_{\mu\mu}}\frac{N_{\Ksmumu}^\text{TOS}}{N_{\Kspipi}^\text{MB}},  
\end{equation}
$N_{\Kspipi}^\text{MB}$ being the number of \Kspipi decays from the MB trigger and $N_{\Ksmumu}^\text{TOS}$ denoting the number of signal decays from the TOS category. $N_{\Kspipi}^\text{MB}$ is found to be around 1000 for every BDT bin.

The quantities
\begin{equation}
\alpha_\text{TIS} = \frac{\eSelect_{\pi\pi}}{\eSelect_{\mu\mu}}  \frac{1}{\epid_{\mu\mu}} \frac{\BRof\Kspipi}{N_{\Kspipi}^\text{TIS}}
\end{equation}
\noindent and
\begin{equation}
\alpha_\text{TOS} = \frac{\eSelect_{\pi\pi}}{\eSelect_{\mu\mu}}\frac{1}{\epid_{\mu\mu}}  \frac{\emb}{\etos_{\mu\mu}}\frac{\BRof\Kspipi}{N_{\Kspipi}^\text{MB}}
\end{equation}
are called {\it normalisation factors} and are defined for each of the BDT bins.
For a given number $N$ of \Ksmumu signal decays, the corresponding value of \BRof\Ksmumu is then $\alpha \times N$.
Using the value of \BRof\Kspipi from Ref.~\cite{PDG}, the normalisation factors are in the range $[6.6,16.2]\times10^{-8}$ for the TIS category, and $[0.9,7.8]\times10^{-8}$ for the TOS category, depending on the BDT bin.
From the normalisation factors, around $2\times10^{-4}$ ($6\times10^{-5}$) SM candidates are expected per BDT bin for the TOS (TIS) analysis. 

\section{Systematic uncertainties}
\label{sec:systematics}

The quantities considered in the determination of the branching fraction that are affected by systematic uncertainties are listed below.
\begin{itemize}
\item The background expectations per bin, obtained by comparing the results with 
the model described in \secref{sec:backgrounds} to those computed: a) if the combinatorial background
is modelled by a linear function;  b) if the mass range over which the fit is performed is modified; 
c) repeating the fit excluding (together with the signal region) the 12\mevcc
left and right windows neighbouring the search window and comparing the fit prediction to the yields in
those regions; no correlation is considered among the different bins for this systematic uncertainty.
\item The ratios of reconstruction and selection efficiencies and absolute muon identification
efficiencies, for which systematic uncertainties are obtained from the difference between different
methods in the data reweighting of the MC computed ratios
and from the comparison to simulation respectively (around 20\% for the ratios and 5\% for muon identification efficiencies); 
no correlation is considered among the different bins.
\item The branching fraction of the normalisation
channel \BRof\Kspipi$= (69.20\pm0.05)\%$~\cite{PDG}; its uncertainty affects
coherently the signal expectations of the forty bins of the analysis.
\item The absolute TOS efficiency,
for which the systematic uncertainty is obtained from the comparison to simulation (around 15\%, depending on the BDT bin); no correlation is considered among the different bins.
\item The effective prescale factor of the MB sample, $\emb = (2.70\pm0.76)\times10^{-6}$.
The uncertainty is evaluated from the difference between the prescale factor as measured in data and the value of the prescale as set in the trigger system.
This  systematic uncertainty  affects coherently the signal expectations of the twenty bins of the TOS analysis.
\end{itemize}

The leading systematic uncertainties are those coming from the absolute TOS efficiency and \emb factor for the TOS analysis and from the ratio of reconstruction and selection efficiencies for the TIS analysis.

\section{Results}
\label{sec:results}

The modified frequentist approach (or \CLs method)~\cite{Read_02,Junk_99} is used to assess the compatibility of the observation with expectations as a function of \BRof\Ksmumu.

Test statistics are built from pseudo-experiments for the signal plus background and background-only hypotheses.
For each pseudo-experiment a product of likelihood ratios is computed depending on the expected number of signal events for a given branching fraction, $s_i$, the expected number of background events, $b_i$ and the observed number of events, $d_i$ for bin $i$.
The \CLsb (\CLb) is defined as the probability for signal plus background (background only) generated pseudo-experiments to have a test-statistic value larger than or equal to that observed in the data.
The \CLs is defined as the ratio of confidence levels $\frac{\CLsb}{\CLb}$. 
This ratio is used to set the exclusion (upper) limit on the branching fraction, whereas $1-\CLb$ is used as a $p$-value to claim evidence or observation. 
A $95 (90)\%$ confidence level exclusion corresponds to $\CLs=0.05 (0.1)$.

The values of $b_i$ are obtained from the fit of the  mass sidebands, as detailed in \secref{sec:backgrounds}.
The values of $s_i$ depend on the assumed branching fraction, as well as on the normalisation factors computed in \secref{sec:normalisation}.
The uncertainties on the input parameters are taken into account by fluctuating the signal and background expectations when generating
the $b$ and $s+b$ ensembles. These fluctuations are performed via asymmetric Gaussian priors, following the formula
\begin{equation}
x'_i = x_i \left(1+ \frac{1}{2}r(s_{+}-s_{-}) + \frac{1}{2}r^2(s_{+}+s_{-})\right) 
\end{equation}
where $x_i$ is the central value of the parameter, $r$ is a random number generated from a normal distribution 
and $s_{+}$ and $s_{-}$ are the relative (signed) errors of $x_i$~\cite{Barlow}. Correlations are implemented by using the same value of $r$ for the parameters that should fluctuate coherently.

\begin{figure}[ht]
  \centering
  \includegraphics[width=0.495\textwidth]{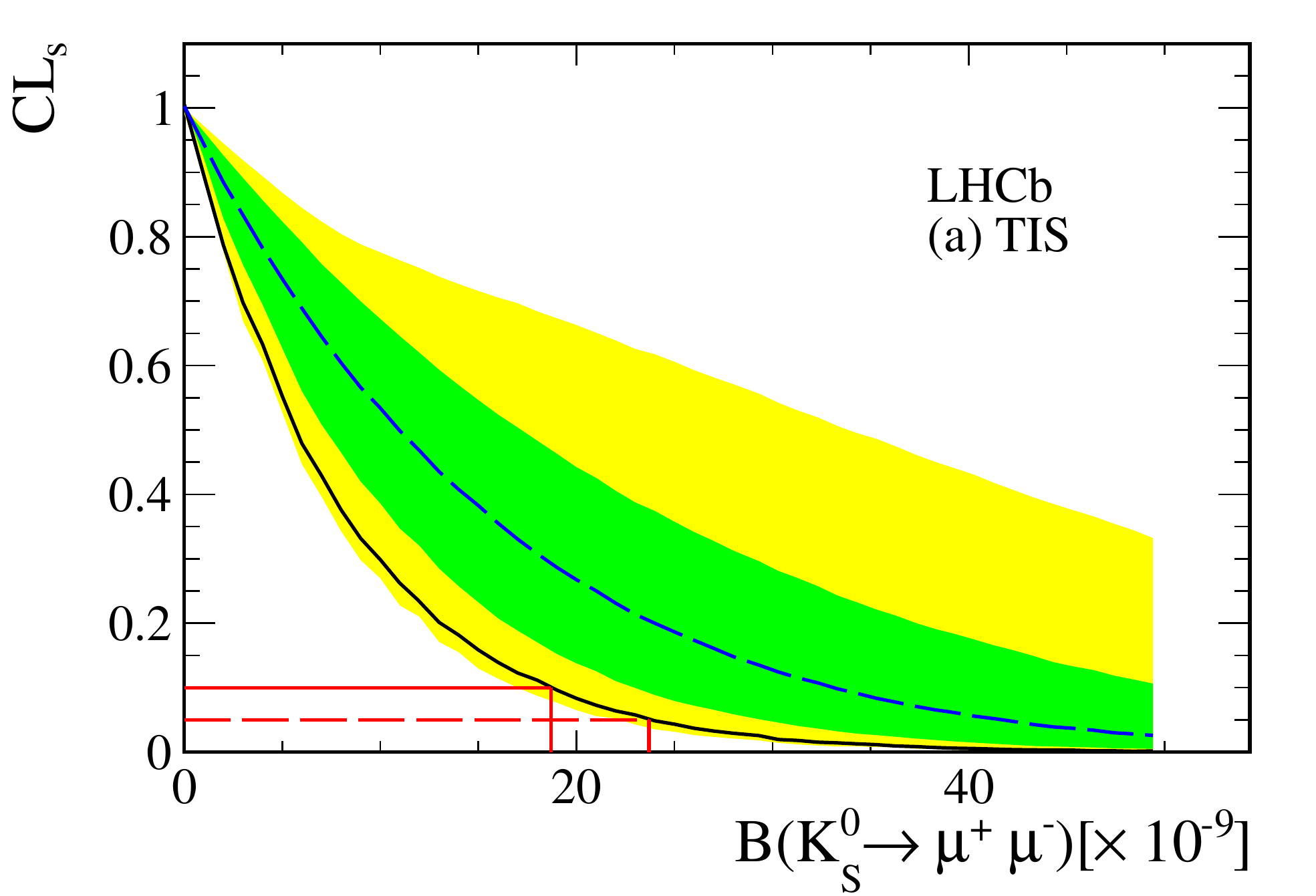}
  \includegraphics[width=0.495\textwidth]{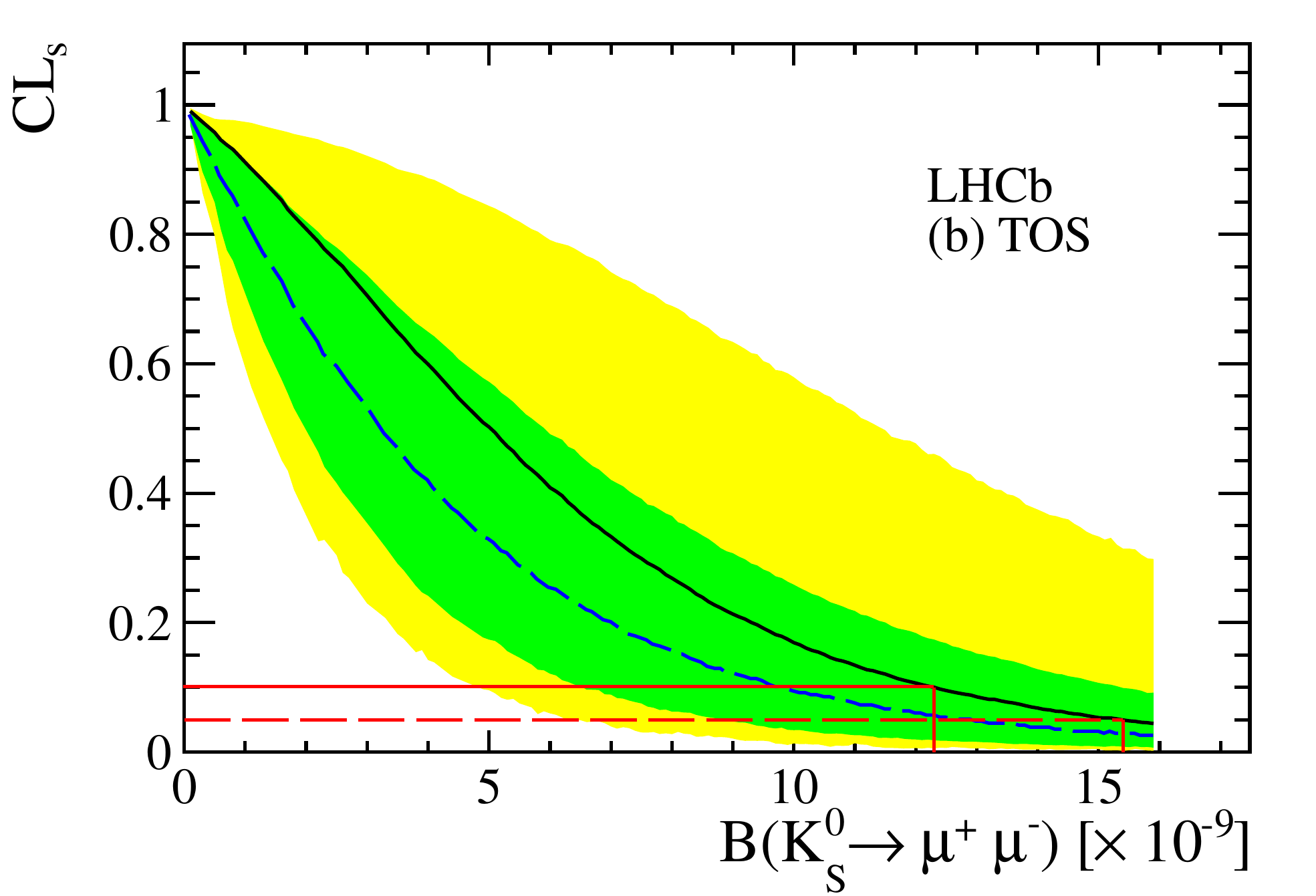}
  \includegraphics[width=0.6\textwidth]  {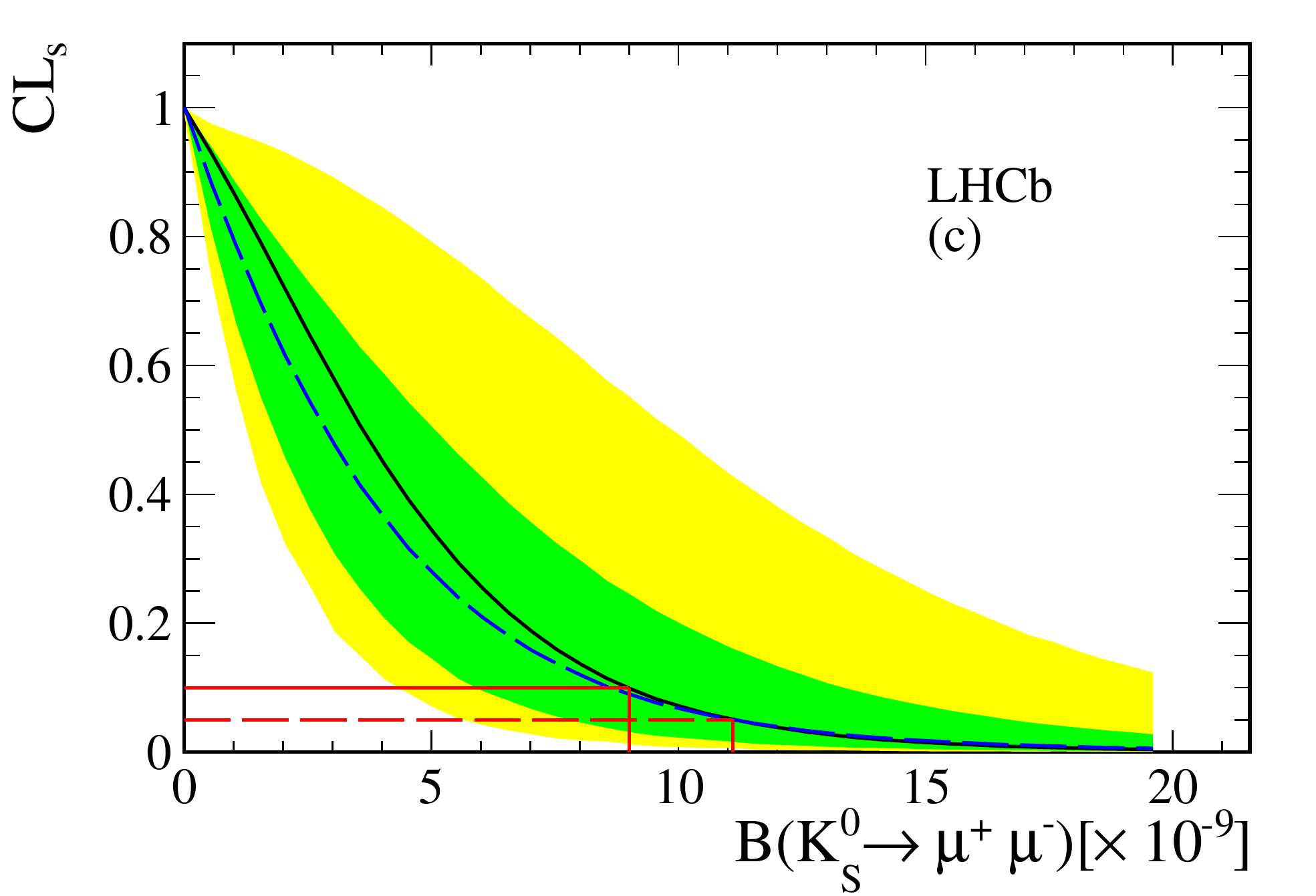}

        \caption{\CLs curves for (a) TIS, (b) TOS categories and for (c) the combined sample. The solid line corresponds to the observed
\CLs. The dashed line corresponds to the median of the \CLs for an ensemble of background-alone experiments. 
In each plot, two bands are shown. The green (dark) band covers $68\%$ ($1\sigma$) of the \CLs curves obtained in the background only pseudo-experiments, while the yellow (light) band covers $95\%$ ($2\sigma$). }
\label{fig:limits}
\end{figure}

\begin{table}[t]
\begin{center}
\tabcaption{Upper limits on \BRof\Ksmumu for the TIS and the TOS categories separately, and for the combined analysis. 
The last entry in the table is the $p$-value of the background-only hypothesis.}
\begin{tabular}{@{}cccc@{}}
\toprule
Quantity & TIS & TOS & Combined \\
\midrule
Expected upper limit at 95\,(90)\% \CL $[10^{-9}]$& $42\,(33)$  & $13\,(10) $ & $11\,(9)$ \\
Observed upper limit at 95\,(90)\% \CL $[10^{-9}]$& $24\,(19)$  & $15\,(12)$ & $11\,(9)$ \\
$p$-value & 0.95 & 0.20 & $0.27$  \\ 
\bottomrule
\end{tabular}
\label{tab:limits}
\end{center}
\end{table}

The observed distribution of events is compatible with background expectations, giving a $p$-value of $27\%$. In particular, in the last 4 bins of the BDT output, corresponding to the most significant region of the analysis, just one candidate is observed in each of the trigger categories, in agreement with the background expectations. 
Figure~\ref{fig:limits} shows the expected and observed \CLs curves for the TIS category and for the TOS category as well as for the combined measurement. 
The upper limit found is 11 (9)$\times 10^{-9}$ at 95 (90)\% confidence level and is a factor of thirty below the previous world best limit.
\tabref{tab:limits} summarises the limits in the TIS, TOS categories, and the combined result.

\section{Conclusions}
\label{sec:conclusions}

A search for \Ksmumu has been performed using 1.0\,fb$^{-1}$ of data collected at the LHCb experiment in 2011.
This search profits from the $10^{13}$ \Ks produced inside the LHCb acceptance and the powerful discrimination against the \Kspipi decay
in which both pions are misidentified as muons, achieved thanks to the LHCb mass resolution for two body decays of the \Ks meson.
The candidates observed are consistent with the expected background, with the $p$-value for the background only hypothesis being $27\%$. The measured upper limit
\begin{equation*} 
\BRof\Ksmumu < 11 (9)\times 10^{-9}
\end{equation*}
\noindent at 95(90)\% confidence level is an improvement of a factor of thirty below the previous world best limit\cite{Jack}.

\section*{Acknowledgements}

\noindent We express our gratitude to our colleagues in the CERN accelerator
departments for the excellent performance of the LHC. We thank the
technical and administrative staff at CERN and at the LHCb institutes,
and acknowledge support from the National Agencies: CAPES, CNPq,
FAPERJ and FINEP (Brazil); CERN; NSFC (China); CNRS/IN2P3 (France);
BMBF, DFG, HGF and MPG (Germany); SFI (Ireland); INFN (Italy); FOM and
NWO (The Netherlands); SCSR (Poland); ANCS (Romania); MinES of Russia and
Rosatom (Russia); MICINN, XuntaGal and GENCAT (Spain); SNSF and SER
(Switzerland); NAS Ukraine (Ukraine); STFC (United Kingdom); NSF
(USA). We also acknowledge the support received from the ERC under FP7
and the Region Auvergne.

\bibliographystyle{LHCb}
\bibliography{main,LHCb_template}

\end{document}